\documentclass[11pt,amsfonts]{amsart}

\usepackage{fullpage}

\def\endproof{\qed \medskip}
\def\blacksquare{\hbox to .60em {\vrule width .60em height .60em}}

\newtheorem{theorem}{Theorem}[section]

\newtheorem{proposition}[theorem]{Proposition}

\newtheorem{remark}[theorem]{Remark}

\begin{document}

\title[]{On the structure of asymptotically de Sitter and Anti-de Sitter spaces}

\author[]{Michael T. Anderson}

\thanks{Partially supported by NSF Grant DMS 0305865}

\maketitle

\abstract
We discuss several aspects of the relation between asymptotically AdS and 
asymptotically dS spacetimes including: the continuation between these 
types of spaces, the global stability of asymptotically dS spaces and the 
structure of limits within this class, holographic renormalization, and the 
maximal mass conjecture of Balasubramanian-deBoer-Minic.
\endabstract

\setcounter{section}{0}

\section{Introduction.}
\setcounter{equation}{0}

 This paper deals with several distinct issues on local and global 
aspects of asymptotically de Sitter spaces and their anti-de Sitter or 
hyperbolic counterparts. Besides their intrinsic interest in classical 
general relativity, asymptotically de Sitter spaces arise frequently in 
the context of inflationary models and issues related to the 
cosmic no-hair conjecture [1], [2]. Moreover, they are of current 
interest in attempts to understand a possible dS/CFT correspondence 
[3], [4] analogous to the much more rigorously established AdS/CFT 
correspondence [5]-[7].

 Asymptotically de Sitter (dS) spaces are understood here to be vacuum 
solutions to the Einstein equations with $\Lambda > 0$ which to the 
future (or past), have geometry asymptotically approaching that of pure 
de Sitter space locally. Globally, these spaces may be quite different 
than de Sitter; in particular future space-like infinity ${\mathcal I}^{+}$ 
may have arbitary topology and induced metric. Asymptotically 
anti-de Sitter (AdS) or hyperbolic (AH) spaces are understood in the 
same sense; see \S 2 for the precise definitions.

\medskip

 It has long been known that there is a formal correspondence between 
asymptotically dS spaces and AH spaces, based on viewing each as a 
hyperboloid in a vacuum ($\Lambda  = 0$) spacetime of one higher 
dimension. The continuation from AH to dS takes place across the null 
cone of a given point, as in the usual hyperboloidal decomposition of 
Minkowski space, cf. [8], and [9]-[11] for instance for more recent 
discussion. In \S 2, this formal process is made exact, and gives a 
rigorous form of Wick rotation or continuation between these classes of 
metrics; in addition, some ambiguities in the choice of 
``analytic continuation'' are addressed.

\medskip

 Perhaps the most significant classical result on de Sitter (dS) 
spacetimes is the stability result of Friedrich [12]; in $3+1$ 
dimensions, the class $dS^{\pm}$ of globally hyperbolic dS spacetimes 
which have smooth and compact future and past conformal infinity 
${\mathcal I}^{+}$ and ${\mathcal I}^{-}$, is open in a natural topology. Thus, 
Cauchy data of any given $(M, g) \in  dS^{\pm}$ may be perturbed in a small 
but arbitrary way, giving rise to spacetimes $(M, \widetilde g)$ in $dS^{\pm}$ 
with the same overall global structure as $(M, g)$. One expects such a result 
is also true in higher dimensions, but a proof of this is lacking.

 In \S 3, we describe the structure of the possible limits of spaces in 
$dS^{\pm}$, i.e. elements in the boundary $\partial (dS^{\pm}) = 
\overline{dS^{\pm}} \setminus dS^{\pm}$, at least in $3+1$ dimensions. All 
limits are globally hyperbolic and geodesically complete and can be one 
of three general types; (I) a pair of spaces $(M, g^{+})$ and $(M, g^{-})$, 
infinitely far apart, and with fully degenerate ${\mathcal I}^{-}$ or ${\mathcal I}^{+}$ 
respectively, or (II) a single space $(M, g^{+})$ with partially or fully 
degenerate ${\mathcal I}^{-},$ or (III), a single space $(M, g^{-})$ with 
partially or fully degenerate ${\mathcal I}^{+}$. In particular, singularities 
occur only for Cauchy data ``outside'' the boundary $\partial (dS^{\pm})$. 
This result is also valid for all dimensions in which Friedrich's theorem holds. 

 From several perspectives, the most natural limits are those of type I. 
As discussed in \S 4, such limit behavior occurs very clearly and explicitly 
in the family of dS Taub-NUT metrics on ${\mathbb R}\times S^{3}$, (and its 
higher dimensional generalizations). No examples are known where the limits 
are of type II or III. 

\medskip

 In \S 5, we discuss holographic renormalization in the dS context, and 
relate the AH, AdS, and dS holographic stress-energy tensors and 
conserved quantities, cf. [13]-[15], [9], [11]. In [15], an interesting 
maximal mass conjecture was proposed for spaces in $dS^{+}$, in analogy 
to positive mass theorems in AdS and AH spaces, and also in analogy to 
entropy bounds in $dS^{+}$ spaces, [16]. Thus, it was conjectured that 
any space $(M, g)$ in $dS^{+}$ has holographic mass $m$ satisfying
\begin{equation} \label{e1.1}
m \leq  m_{0},
\end{equation}
where $m_{0}$ is the mass of pure de Sitter space (in static 
coordinates), unless $(M, g)$ has a cosmological singularity. In 
particular, any space $(M, g) \in dS^{\pm}$ should satisfy (1.1). We 
refer to [17]-[20] for prior work and commentary on this conjecture.

 However, we find in \S 5 general counterexamples to this conjecture in 
$3+1$ dimensions. Depending on the exact formulation of the conjecture, 
counterexamples are also found in $n+1$ dimensions, for all $n$ odd, 
and there is strong evidence that it fails also for all $n$ even.

\medskip

  I would like to thank V. Balasubramanian, J. de Boer and D. Minic 
as well as R. Mann and A. Ghezelbash for interesting correspondence 
on these issues.

\section{Asymptotic Analytic Continuation.}
\setcounter{equation}{0}

 Let $M$ be the interior of a compact manifold $\bar M$ with 
boundary $\partial M$. A complete Riemannian metric $g$ on $M$ is 
conformally compact if there is a defining function $\rho : \bar M  
\rightarrow  {\mathbb R}$ such that the compactified metric 
\begin{equation} \label{e2.1}
\bar g = \rho^{2}g 
\end{equation}
extends smoothly to a Riemannian metric on $\partial M.$ A defining 
function $\rho$ satisfies $\rho \geq 0$, $\rho^{-1}(0) = \partial M$ 
and $d\rho  \neq 0$ on $\partial M.$ The induced metric $\gamma  = 
\bar g|_{\partial M}$ is called a boundary metric for $g$. Since 
defining functions are unique only up to multiplication by positive 
functions, only the conformal class $[\gamma]$ is an invariant of 
$(M, g)$ and $[\gamma]$ is called the conformal infinity of $(M, g)$. 
we will always assume that $M$ is connected, although the boundary 
$\partial M$ may be connected or disconnected.  

 If $g$ is an Einstein metric on $M$, so that 
\begin{equation} \label{e2.2}
Ric_{g} - \frac{R}{2}g +\Lambda g = 0, 
\end{equation}
then it is easy to see that if $\Lambda  > 0$, or $\Lambda  = 0$, then 
$(M, g)$ cannot be conformally compact; in the former case $\partial M = 
\emptyset$ while in the latter case one has at best $\partial M = \{pt\}$. 
Thus, $\Lambda  < 0$ and such metrics are called asymptotically hyperbolic 
(AH) or asymptotically Euclidean anti de Sitter (EAdS) Einstein metrics. 
As usual, define the length scale $l$ by $\Lambda  = -n(n-1)/2l^{2}$. 

 A compactification $\bar g = \rho^{2}g$ as in (2.1) is geodesic 
if $\rho (x) = dist_{\bar g}(x, \partial M)$. Such 
compactifications are especially useful for computational purposes, and 
for the remainder of the paper we work only with geodesic 
compactifications. Each choice of boundary metric $\gamma \in [\gamma]$ 
determines a unique geodesic defining function $\rho$. If $l \neq 1$, 
it is convenient to work with the dimensionless function $\rho  = 
dist_{\bar g}(\partial M, \cdot  )/l$. 

 The Gauss Lemma gives the splitting
\begin{equation} \label{e2.3}
\bar g = l^{2}(d\rho^{2} + g_{\rho}), \ \ 
g = (\frac{l}{\rho})^{2}(d\rho^{2} + g_{\rho}), 
\end{equation}
where $g_{\rho}$ is a curve of metrics on $\partial M$. In the 
following, it is convenient to work in the scale $l = 1$, (which 
can always be achieved by rescaling (2.3). The Ricatti 
equation associated to the splitting (2.3) then states
$$A'  + A^{2} + R_{T} = 0, $$
where $A$ is the 2nd fundamental form of the level sets $S(\rho)$ of 
$\rho$, $A'  = \nabla_{T}A$ with $T = \nabla\rho$ and $R_{T}(V) = 
\langle R(T,V)V,T \rangle$; all computations here are with respect to 
$\bar g$. Using this and the Gauss equations for $S(\rho) \subset  
(M, \bar g)$ together with standard formulas for the curvatures of 
conformally equivalent metrics, one obtains for geodesic 
compactifications the relation
\begin{equation} \label{e2.4}
\rho\ddot g_{\rho} -(n-1)\dot g_{\rho} - 2Hg_{\rho} = 
\rho [2Ric_{\rho} - H\dot g_{\rho} + (\dot g_{\rho})^{2} - 
2(Ric_{g} + n|d\rho|^{2}g)^{T}],
\end{equation}
where $\dot g_{\rho} = {\mathcal L}_{T}g_{\rho} = \frac{1}{2}A$, $Ric_{\rho}$ 
is the intrinsic Ricci curvature of the level sets $S(\rho)$ of $\rho$, 
$H = tr A$ is the mean curvature of $S(\rho)$ in $(M, g)$ and $T$ denotes 
orthogonal projection onto the level sets $S(\rho)$. 

 In particular, when $g$ is Einstein, $Ric_{g} + n|d\rho|^{2}g = 0$ and 
(2.4) gives
\begin{eqnarray} \label{e2.5}
\rho\ddot g_{\rho} -(n-1)\dot g_{\rho} - 2Hg_{\rho} = 
\rho [2Ric_{\rho} - H\dot g_{\rho} + (\dot g_{\rho})^{2}], \\  
\rho \dot H - H = \rho |A|^{2}, \nonumber \\
\delta A = -dH, \nonumber
\end{eqnarray}
where the latter two equations arise from the trace of the Ricatti equation 
and the Gauss-Codazzi equations respectively. 

  Setting $g_{0} = g_{(0)} = \gamma$, one has $g_{(1)} = 
\dot g_{\rho}|_{\rho =0} = 0$. By differentiating (2.5) with respect to 
$\rho$ inductively, one obtains a formal series expansion for the 
curve $g_{\rho}$; this is the Fefferman-Graham expansion [21]. The 
exact form of the expansion depends on whether $n$ is odd or even. If 
$n$ is odd, then 
\begin{equation} \label{e2.6}
g_{\rho} \sim  g_{(0)} + \rho^{2}g_{(2)} + .... + \rho^{n-1}g_{(n-1)} + 
\rho^{n}g_{(n)} + \rho^{n+1}g_{(n+1)} + ... 
\end{equation}
This expansion is even in powers of $\rho$ up to order $n-1$. The 
coefficients $g_{(2k)}$, $k \leq (n-1)/2$ are locally determined by the 
boundary metric $\gamma  = g_{(0)}$; they are explicitly computable 
expressions in the curvature of $\gamma$ and its covariant 
derivatives. The last two equations in (2.5) imply that the term $g_{(n)}$ 
is transverse-traceless, i.e.
\begin{equation} \label{e2.7}
tr_{\gamma}g_{(n)} = 0,  \ \ \delta_{\gamma}g_{(n)} = 0, 
\end{equation}
but $g_{(n)}$ is otherwise undetermined by $\gamma$; it depends on the 
global structure of the AH Einstein metric $(M, g)$. For $k > n$, terms 
$g_{(k)}$ occur for both $k$ even and odd; the term $g_{(k)}$ depends 
on two boundary derivatives of $g_{(k-2)}$.

 If $n$ is even, one has
\begin{equation} \label{e2.8}
g_{\rho} \sim  g_{(0)} + \rho^{2}g_{(2)} + .... + \rho^{n-2}g_{(n-2)} + 
\rho^{n}g_{(n)} + \rho^{n}\log\rho \ h_{(n)} + ... 
\end{equation}
Again the terms $g_{(2k)}$ up to order $n-2$ are explicitly computable 
from the boundary metric $\gamma$, as is the transverse-traceless 
coefficient $h_{(n)}$ of the first $\log\rho$ term. The term $h_{(n)}$ 
is an important term for the corresponding CFT on $\partial M$; up to a 
constant, it is the metric variation of the conformal anomaly, cf. [14], 
(or also [22]). The term $g_{(n)}$ satisfies 
\begin{equation} \label{e2.9}
tr_{\gamma}g_{(n)} = \tau ,  \ \ \delta_{\gamma}g_{(n)} = \delta ,  
\end{equation}
where $\tau$ and $\delta$ are explicitly determined by the boundary 
metric $\gamma$ and its derivatives, but again $g_{(n)}$ is otherwise 
undetermined by $\gamma$. The series (2.8) is even in powers of $\rho$, 
and terms of the form $\rho^{k}(\log\rho)^{l}$ appear at order $k > n$. 
Again the coefficients $g_{(k)}$ and $h_{(k)}$ depend on two derivatives 
of $g_{(k-2)}$ and $h_{(k-2)}$. The expansions (2.6) and (2.8) of course 
depend on the choice of boundary metric $\gamma \in [\gamma]$. However, 
transformation properties of the coefficients $g_{(n)}$ and $h_{(n)}$ 
under conformal changes are readily computable, cf. [14], [23].

 Mathematically, the expansions (2.6) and (2.8) are formal, obtained by 
compactifiying the Einstein equations and taking iterated Lie 
derivatives of $\bar g$ at $\rho = 0$. If $\bar g \in  
C^{m,\alpha}(\bar M)$, then the expansions hold up to order 
$m+\alpha$. However, boundary regularity results are needed to ensure 
that if an AH Einstein metric $g$ with boundary metric $\gamma$ 
satisfies $\gamma  \in  C^{m,\alpha}(\partial M)$, then the 
compactification $\bar g \in  C^{m,\alpha}(\bar M)$. These have 
recently been established in general in [24] for $n = 3$, and in [25] 
for $n \geq 3$ in case of $C^{\infty}$ boundary metric $\gamma$. 

\medskip

 In sum, an AH Einstein metric is formally determined near $\partial M$ 
by $g_{(0)}$ and $g_{(n)}$. The term $g_{(0)}$ corresponds to Dirichlet 
boundary data on $\partial M,$ while $g_{(n)}$ corresponds to Neumann 
boundary data, (in analogy with the scalar Laplace operator). Thus, for 
global AH Einstein metrics defined on a compact manifold with boundary 
$\bar M = M \cup \partial M$, the correspondence
\begin{equation} \label{e2.10}
g_{(0)} \rightarrow  g_{(n)} 
\end{equation}
is analogous to the Dirichlet-to-Neumann map for harmonic functions. 
As discussed in \S 5, the term $g_{(n)}$ is essentially the 
(renormalized) Brown-York quasilocal stress-energy tensor of $(M, g)$ and 
corresponds to the expectation value of the stress-energy tensor of the dual 
CFT on $\partial M$ in all known cases. However, the map (2.10) per-se is only 
well-defined if there is a unique AH Einstein metric with boundary data 
$\gamma  = g_{(0)}$. This occurs in some situations, but fails in others.

 On the other hand, the expansions (2.6) and (2.8) are completely 
local, both in the distance to the boundary $\rho$, and tangentially 
on $\partial M$. If one restricts only to local considerations, then 
the free data in the expansion are $g_{(0)}$ and $g_{(n)}$, subject to 
the constraint equations (2.7) and (2.9). Of course arbitrarily given 
$g_{(0)}$ and $g_{(n)}$ will not correspond to globally smooth solutions 
on a compact manifold $\bar M$. 

\medskip

 Next we turn to Lorentzian metrics. The definition of a conformal 
completion for Lorentz metrics is similar, although more subtle, than 
the Riemannian case due to the causal structure and the common 
occurence of singularities. Note however that for local considerations, 
such as the computation of the coefficients $g_{(k)}$ in the expansions 
(2.6) or (2.8), global issues are irrelevant. 

\medskip

 As is well-known, the structure of the boundary of a conformal 
completion depends on the sign of $\Lambda .$ If $\Lambda < 0$, the 
AdS case, then $\partial M = {\mathbb R}\times \partial\Sigma$, with $\gamma$ 
a Lorentz metric. For simplicity, we will always assume that 
$\partial\Sigma$ is compact and, abusing notation slightly, call $(M, g)$ 
conformally compact if the definition (2.1) holds with $\partial M$ 
of the form above. Thus, $\partial\Sigma$ is the conformal boundary of 
a space-like slice $\Sigma$ in $(M, g)$, while $\partial M = {\mathcal I}$ is 
timelike. Note that ${\mathcal I}$ is not necessarily connected, as is the case 
for instance for the AdS Schwarzschild metric.

 If $\Lambda > 0$, the dS case, then $(\partial M, \gamma)$ is a 
Riemannian manifold, representing the spatial behavior at future or 
past infinite times. Thus, $\partial M$ may have two space-like 
components ${\mathcal I}^{+}$, ${\mathcal I}^{-}$ representing future and past conformal 
infinity respectively, or possibly only one component, ${\mathcal I}^{+}$ or 
${\mathcal I}^{-}$. The definition (2.1) holds without change in this situation. 
However, the compactness of $\partial M$ precludes the existence of (most) 
singularities in $(M, g)$, and so many important solutions are not 
conformally compact. We will say that $(M, g)$ is partially conformally 
compact if $\partial M$ is open and non-empty. Note that $\gamma \in 
[\gamma]$ may then either be complete or incomplete.

 If $\Lambda  = 0$, the flat or Minkowski case, then $(\partial M, 
\gamma)$ is null (degenerate), of the form $\partial M = {\mathbb R} 
\times \partial\Sigma$. Usually $\partial \Sigma$ is taken to be 
compact, and represents the boundary at infinity of a null-hypersurface in 
$(M, g)$. The ${\mathbb R}$ factor parametrizes null geodesics, and again 
$\partial M$ may have two components, ${\mathcal I}^{+}, {\mathcal I}^{-}$.

\medskip

  Consider now the expansions (2.6) or (2.8) in these situations.

 I. Suppose $\Lambda < 0$, the AdS case. Then $(M, g)$ is spatially 
non-compact and conformal infinity $(\partial M, \gamma)$ is 
Lorentzian; the vector field $\partial_{\rho} = \nabla\rho$ is 
space-like. All of the discussion (2.3)-(2.9) above holds without change 
in this setting; the only difference is that $g_{\rho}$ is curve of Lorentz 
metrics on $S(\rho)$. 

 II. Suppose $\Lambda > 0$, the dS case. Then conformal infinity 
$(\partial M, \gamma)$ has components ${\mathcal I}^{+}$, ${\mathcal I}^{-}$ 
or both. There is no geodesic defining function $\rho$ for both 
${\mathcal I}^{+}$ and ${\mathcal I}^{-}$ simultaneously; instead one has 
$\rho  = \rho^{+}$ or $\rho  = \rho^{-}$. In both situations, $\partial_{\rho}$ 
is time-like, and $\partial_{\rho} = -\nabla\rho$. Thus $\partial_{\rho}$ 
is past directed near ${\mathcal I}^{+}$ and future directed near ${\mathcal I}^{-}$. 

 Since $|d\rho|^{2} = -1$, (or $-l^{-2}$), (2.5) changes to its 
negative, i.e.
\begin{equation} \label{e2.11}
\rho\ddot g_{\rho} -(n-1)\dot g_{\rho} -2Hg_{\rho} = 
-\rho [2Ric_{\rho} - H\dot g_{\rho} + (\dot g_{\rho})^{2}]. 
\end{equation}
This leads to an expansion of the form (2.6) or (2.8) exactly as in the 
AH or EAdS case; for $n = 3$ this expansion is due to Starobinsky [26], 
predating the work of Fefferman-Graham. To compare these expansions, suppose 
\begin{equation} \label{e2.12}
g_{(0)}^{dS} = g_{(0)}^{AH}. 
\end{equation}
Then (2.5) and (2.12) immediately give $g_{(1)}^{dS} = g_{(1)}^{AH} = 
0$, while $g_{(2)}^{dS} = -g_{(2)}^{AH}$, (when $n > 2$). Similarly, for 
$2k < n$, 
\begin{equation} \label{e2.13}
g_{(2k)}^{dS} = (-1)^{k}g_{(2k)}^{AH}. 
\end{equation}
Recall that the transverse-traceless part of the term $g_{(n)}$ is 
undetermined; this occurs on both the AH and dS sides. The natural 
relation between $g_{(n)}^{AH}$ and $g_{(n)}^{dS}$ is to set
\begin{equation} \label{e2.14}
g_{(n)}^{dS} = \pm g_{(n)}^{AH}. 
\end{equation}
However, there appears to be no compelling reason to choose one sign 
over the other. The continuation of pure hyperbolic space to pure 
de Sitter space, (in static coordinates), requires
\begin{equation} \label{e2.15}
g_{(2)}^{dS} = -g_{(2)}^{AH}, \ {\rm for} \ n = 2, \ \ {\rm and} \ \ 
g_{(4)}^{dS} = g_{(4)}^{AH}, \ {\rm for} \ n = 4, 
\end{equation}
but imposes no restrictions for other $n$. Based on the examples 
discussed in \S 4, we choose
\begin{equation} \label{e2.16}
g_{(n)}^{dS} = \pm g_{(n)}^{AH}, 
\end{equation}
where $+$ is chosen if $n \equiv  0,1$ (mod 4), while $-$ is chosen if $n 
\equiv 2,3$ (mod 4); see also Remark 2.2 for the opposite choice. The 
coefficients of the AH and dS expansions are then related by 
\begin{equation} \label{e2.17}
g_{(k)}^{dS} = \pm g_{(k)}^{AH} \ \ {\rm and} \ \ h_{(n)}^{dS} = 
(-1)^{n/2} h_{(n)}^{AH},
\end{equation}
for $k > n$, where again $+$ occurs if $k \equiv 0,1$ (mod 4), while $-$ 
occurs if $k \equiv 2,3$ (mod 4); see the Appendix for further 
details. 

 Thus, one has a formal correspondence between AH and dS solutions of 
the Einstein equations, given by 
\begin{equation} \label{e2.18}
g^{AH} = (\frac{l}{\rho})^{2}(d\rho^{2} + g_{\rho}^{AH}) 
\Longleftrightarrow g^{dS} = (\frac{l}{\rho})^{2}(-d\rho^{2} + 
g_{\rho}^{dS}), 
\end{equation}
where the formal expansions of $g_{\rho}^{AH}$ and $g_{\rho}^{dS}$ have 
coefficients related by (2.17). The metrics $g^{AH}$ and $g^{dS}$ can 
be formally obtained from each other by changing $\rho$ to $i\rho$ 
and $l$ to $il$ in $g^{AH}$, and dropping all resulting $i$ coefficients 
(at odd powers of $\rho$) in the expansions for $g_{\rho}^{dS}$. Note that 
a given AH metric generates a de Sitter metric $g^{dS}$ with either a future 
conformal infinity ${\mathcal I}^{+}$ or a past conformal infinity ${\mathcal I}^{-}$, 
but not necessarily both simultaneously.

 To actually construct or prove the existence of metrics from this correspondence, 
one needs the formal expansions to converge to $g_{\rho}$.
\begin{theorem} \label{t 2.1.}
There is a 1-1 correspondence between $C^{\omega}$ conformally compact 
Riemannian AH Einstein metrics with boundary data $(\gamma, g_{(n)})$, and 
$C^{\omega}$ conformally compact Lorentzian dS Einstein metrics with past (or 
future) boundary data $(\gamma, g_{(n)})$, given by (2.18). Thus, given any 
real-analytic metric $\gamma$ and any real-analytic symmetric bilinear form 
$g_{(n)}$ on an $n$-manifold $\partial M$, satisfying the constraint conditions 
(2.7) or (2.9), there exist unique AH Einstein and dS Einstein metrics with 
boundary data $(\gamma, g_{(n)})$, defined in a thickening of $\partial M$, 
and related by the correspondence (2.18).

\end{theorem}
\noindent
{\bf Proof:}
 When $n = 3$, the existence and convergence of the expansions on the 
AH and dS sides has been proved in [24], and the discussion above then 
gives the correspondence (2.18). For general $n$, the existence and 
convergence of the expansions has recently been proved in [27] and [28]. 
{\endproof}

 This correspondence thus gives a rigorous form of ``Wick rotation'' 
between these types of Einstein metrics. The construction can be 
realized geometrically by embedding the spaces $(M, g^{AH})$ and $(M, 
g^{dS})$ into a self-similar vacuum spacetime ($\Lambda  = 0$) of one 
higher dimension, exactly as the embeddings of the hyperbolic and 
de Sitter metrics as hyperboloids in flat Minkowski space. The 
transition from AH geometry to dS geometry takes place across the 
future or the past light cone of a fixed point.

 However, this construction is not, strictly speaking, an analytic 
continuation of AH metrics to dS metrics. When $n$ is odd, the 
compactified metrics $g^{AH}$ and $g^{dS}$ are only $C^{n/2}$ across 
conformal infinity, in the generic case when $g_{(n)} \neq 0$. If $n$ 
is even and the conformal anomaly stress-energy $h_{(n)}$ is non-vanishing, 
the compactifcations $g^{AH}$ and $g^{dS}$ are $C^{n/2-\varepsilon}$ 
across conformal infinity.

\medskip

 The Einstein equations on the AH side form an elliptic system of 
PDE's, in a suitable gauge. Thus, at least formally, one has a 
well-defined Dirichlet problem at infinity. For global solutions defined 
on a compact manifold with boundary $\bar M = M \cup \partial M$, the 
stress-energy term $g_{(n)}$ is thus determined (formally) by the boundary 
metric $\gamma  = g_{(0)}$. On the dS side, the Einstein equations 
form, again in suitable gauge, a hyperbolic system of PDE. Here the 
terms $g_{(0)}$ and $g_{(n)}$ form initial or Cauchy data for the 
evolution equations. Thus, they are freely prescribed, subject to the 
constraints (2.7) or (2.9) and independent of each other; of course 
arbitrary data $g_{(0)}$, $g_{(n)}$ may not give rise to global solutions.

\medskip

 This distinction is reflected in the behavior of symmetries of the 
boundary metrics at conformal infinity. If $G$ is a connected group of 
conformal isometries of $(\partial M, [\gamma])$, then $G$ extends to 
a group of isometries of any AH Einstein filling metric $(M, g)$ with 
boundary data $(\partial M, [\gamma])$, cf. [29]. This is of course not 
the case on the dS side. Symmetries of $\gamma$ do not necessarily extend 
to symmetries of $g^{dS}$; only isometry groups preserving both $\gamma$ 
and $g_{(n)}$ extend to isometries of $g^{dS}$.
\begin{remark} \label{r 2.2.}
  {\rm As noted above in the proof of Theorem 2.1, when $n$ is odd and 
$\gamma$ is real-analytic, the series (2.6) converges. Hence, one can 
continue an AH Einstein metric $(M, g)$ to the region $\rho < 0$, 
obtaining an AH Einstein metric $\widetilde g$ ``on the other side'' 
of $\partial M$. One then has 
\begin{equation} \label{e2.19}
\widetilde g_{(0)} = g_{(0)} \ \ {\rm and} \ \ \widetilde g_{(n)} = 
-g_{(n)}. 
\end{equation}
For example, when $n = 3$, and for $\rho > 0$, the expansion for 
$\widetilde g$ has the form
\begin{equation} \label{e2.20}
\widetilde g_{\rho} = g_{(0)} + \rho^{2}g_{(2)} - \rho^{3}g_{(3)} + 
\rho^{4}g_{(4)} - \rho^{5}g_{(5)} + ... ,
\end{equation}
with $\rho > 0$. This has the dS continuation
\begin{equation} \label{e2.21}
\widetilde g_{\rho}^{dS} = g_{(0)} - \rho^{2}g_{(2)} + \rho^{3}g_{(3)} + 
\rho^{4}g_{(4)} - \rho^{5}g_{(5)} + ... 
\end{equation}
Here, and for all $n$ odd, one thus has the opposite signs from (2.17), 
for $k$ odd; the series for $\widetilde g_{\rho}^{dS}$ is obtained from that of 
$g_{\rho}^{AH}$ by replacing $\rho$ by $-i\rho$, (instead of $\rho  
\rightarrow  i\rho$).

 However, in all known examples, if $(M, g)$ is a globally smooth AH 
Einstein metric, the continuation $\widetilde g$ has singularities; for 
example the positive mass AH Schwarzschild metrics continue in $\rho < 0$ 
to the corresponding negative mass AH Schwarzschild metrics, cf. 
also [30] for other examples. 

 When $n$ is even, the term $\log \rho$ cannot be continued to $\rho  
< 0$, and so one does not obtain an analogous $\widetilde g$ in this 
case. This also suggests that (2.16) is the right sign choice for $n$ 
even. }
\end{remark}

\begin{remark} \label{r 2.3.}
  {\rm Theorem 2.1 also holds for the continuation of AdS, (i.e Lorentzian) 
Einstein metrics. However, the continuation is then a solution of the 
Einstein equations with $\Lambda > 0$, and with signature $(2, n-1)$, i.e. 
$(--++ ... +)$. In some circumstances, this can be Wick rotated to a 
Lorentzian metric, just as an AdS metric can sometimes be Wick rotated 
to an AH metric. }
\end{remark}

\begin{remark} \label{r 2.4.}
  {\rm The analogue of the expansion (2.6) or (2.8) when $\Lambda = 0$ is 
much more complicated, and goes back to work of Bondi and Sachs; cf. 
[31] for a recent discussion. }
\end{remark}

\section{Global structure of dS spaces.}
\setcounter{equation}{0}

 Let $dS^{+}$ be the space of de Sitter spacetimes, i.e. vacuum 
solutions of the Einstein equations with $\Lambda  >$ 0, which are 
conformally compact to the future; thus $({\mathcal I}^{+}, \gamma^{+})$ is 
a compact Riemannian manifold, without boundary, with $\gamma^{+} \in  
[\gamma^{+}].$ The same definition holds for $dS^{-}.$ Let $dS^{\pm}$ 
be the space of de Sitter spacetimes which are globally conformally 
compact. If $(M, g) \in  dS^{\pm}$ is globally hyperbolic, then it is 
easy to see that $M$ is geodesically complete and topologically a 
product of the form $M = {\mathbb R} \times \Sigma$, with Cauchy surface 
$\Sigma $ diffeomorphic to ${\mathcal I}^{+}$ and ${\mathcal I}^{-}$. It will 
always be assumed that spaces in $dS^{\pm}$ are globally hyperbolic.
\begin{remark} \label{r 3.1.}
 {\rm  A spacetime $(M, g)$ in $dS^{+}\cap dS^{-}$ is necessarily globally 
hyperbolic in a neighborhood of ${\mathcal I}^{+}$ and ${\mathcal I}^{-},$ and 
it is natural to conjecture that all of $(M, g)$ is globally hyperbolic. 
However, this is an open problem. Apriori, there may be singularities 
sandwiched between the Cauchy surfaces near ${\mathcal I}^{+}$ and ${\mathcal I}^{-}$ 
which don't propagate to either ${\mathcal I}^{+}$ or ${\mathcal I}^{-}$. Although 
physically unlikely to occur, mathematically such solutions have not been 
ruled out, cf. [32] for further discussion. }
\end{remark}
 For $(M, g) \in  dS^{+}$, one has the data at infinity $(g_{(0)}^{+}, 
g_{(n)}^{+})$ on ${\mathcal I}^{+}$, from the Fefferman-Graham or Starobinsky 
asymptotic expansion. At least for analytic data, and up to a natural 
equivalence, these uniquely determine, up to isometry, the maximal 
globally hyperbolic solution $(M, g)$. Thus, let
\begin{equation} \label{e3.1}
{\mathcal C}^{+} \subset  Met({\mathcal I}^{+})\times S^{2}({\mathcal I}^{+}), 
\end{equation}
be the subset consisting of pairs $(\gamma, g_{(n)})$ such that (2.7) 
or (2.9) holds, depending on whether $n$ is odd or even. Let ${\mathcal A}^{+} 
= {\mathcal C}^{+}/ \sim$, where $\gamma_{1} \sim \gamma_{2}$ if there exists 
$\sigma:{\mathcal I}^{+} \rightarrow {\mathbb R}$ such that $\gamma_{2} = 
e^{2\sigma}\gamma_{1}$, and $(g_{1})_{(n)} \sim (g_{2})_{(n)}$ 
if $(g_{2})_{(n)} = e^{-n\sigma}(g_{1})_{(n)}$ when $n$ is odd; see [14], 
[23] for the more complicated relation on $g_{(n)}$ if $n$ is even.

 Then in any dimension, any real-analytic data $(\gamma , g_{(n)}) \in  
{\mathcal A}^{+}$ determine a unique solution up to isometry in $dS^{+}$, by 
[27] or [28]. When $n = 3$, so that the spacetime is 4-dimensional, a result 
of Friedrich [12] shows that $C^{k}$ data in ${\mathcal A}^{+}$ determine a unique 
solution in $dS^{+}$, for $k \geq 7$ for instance. Of course the same results 
hold with ${\mathcal C}^{-}$ and $dS^{-}$.

\medskip

 When $n = 3$, a basic result of Friedrich [12] is that $dS^{\pm}$ is 
open in the $C^{k}$ topology, i.e. given any solution $(M, g_{0}) \in  
dS^{\pm}$, any sufficiently small perturbation of the initial data in 
${\mathcal A}^{-}$ at ${\mathcal I}^{-}$, (or of the final data in ${\mathcal A}^{+}$ 
at ${\mathcal I}^{+}$) gives rise to a global solution $(M, g)$ in $dS^{\pm}$ 
near $(M, g_{0})$. This applies in particular to de Sitter space itself, 
so any small perturbation of the pure dS data $(\gamma_{S^{3}(1)}, 0)$ on 
${\mathcal I}^{+}$ or ${\mathcal I}^{-}$ gives rise to a complete solution 
$(M, g)$ in $dS^{\pm}$.

 One certainly expects Friedrich's theorem to be true in all 
dimensions, but a rigorous mathematical proof of this is lacking. It 
would suffice to prove a Cauchy stability theorem for the degenerate 
(Fuchsian) system of PDE obtained by conformally compactifying the 
Einstein equations, as in (2.5). For analytic boundary data 
say at ${\mathcal I}^{+},$ Fuchsian versions of the Cauchy-Kovalevsky 
theorem give the existence of the analytic or polyhomogenous solutions 
$(M, g) \in  dS^{+}$ mentioned above. However, it is well-known that in 
general, solutions given by the Cauchy-Kovalevsky theorem do not 
necessarily vary continuously with the initial data.

\medskip

 In the opposite direction, there are two general results, valid in all 
dimensions, restricting the boundary data on ${\mathcal I}^{+}$ or ${\mathcal 
I}^{-}$ of metrics in $dS^{\pm}.$ First, it is proved in [33], [34] 
that if $(M, g) \in  dS^{+}$ has a representative metric $\gamma^{+}\in 
 [\gamma^{+}]$ of negative scalar curvature, $R_{\gamma^{+}} < $ 0, (or 
$R_{\gamma^{+}} \leq $ 0), then ${\mathcal I}^{-} = \emptyset  ,$ so that 
there is not even a partial compactification at past infinity. In fact 
if $\gamma^{+}$ is chosen to have constant negative scalar curvature $R 
<$ 0, there is an upper bound $\rho_{0}$ on the distance to ${\mathcal 
I}^{+}$ in the geodesic compactification,
\begin{equation} \label{e3.2}
\rho  \leq  \rho_{0} = \frac{4n(n-1)}{|R|}. 
\end{equation}

 Second, the following interesting result has been proved in [33]; if 
$(M, g) \in  dS^{+}$ and $|\pi_{1}({\mathcal I}^{+})| = \infty$, then 
again ${\mathcal I}^{-} = \emptyset$. When $n = 3$, the only known 
3-manifolds with finite fundamental group are spherical space-forms 
$S^{3}/\Gamma$. Perelman's work [35] implies that in fact these are 
the only such 3-manifolds. Up to finite covering projections, it 
follows that one must have ${\mathcal I}^{+} = S^{3}$ and $M = {\mathbb R} 
\times S^{3}$ for spaces in $dS^{\pm}$, regardless of the data in 
${\mathcal C}^{+}$.

\medskip

 Consider the (gravitational) scattering map 
\begin{equation} \label{e3.3}
{\mathcal S} : {\mathcal A}^{-} \rightarrow  {\mathcal A}^{+}, 
\end{equation}
$${\mathcal S} [\gamma^{-}, g_{(n)}^{-}] = [\gamma^{+}, g_{(n)}^{+}]. $$
Here $g_{(n)}^{\pm}$ are taken with respect to the future normal 
direction; hence $g_{(n)}^{+}$ differs in sign from the term $g_{(n)}$ 
in (2.6) or (2.8), since at ${\mathcal I}^{+}$, $\partial_{\rho}$ is past 
directed.

 When $n = 3$, using either the linearized conformal field equations [12] 
or the linearized Bach equations, cf. [25], on the compactified 
(unphysical) metric $\bar g,$ it is straightforward to prove that 
$D_{g}{\mathcal S}$ is an isomorphism, for any $g\in dS^{\pm}$. This is 
because the linearized equations at $\bar g$ are a linear system 
of wave equations, (in a suitable gauge), on a compact manifold with 
finite time interval. Again, one expects such a result to be true in 
all dimensions, but this remains to be proved.

 For $n = 3$, it follows that one may locally parametrize $dS^{\pm}$ by 
conformal classes of metrics $(\gamma^{-}, \gamma^{+})$ on ${\mathcal I}^{-}
\times {\mathcal I}^{+}$ instead of ${\mathcal A}^{-}$ or ${\mathcal A}^{+}$. 
Further, $||D{\mathcal S}_{g}||$ is bounded, with a bound depending on $g$. 
However, as will be seen below and in \S 4, $||D_{g}{\mathcal S}||$ blows up on 
bounded sequences of data in ${\mathcal A}^{-}$ or ${\mathcal A}^{+}$, or 
${\mathcal I}^{-}\times {\mathcal I}^{+}$.

\medskip

 It is of interest to examine the structure of the closure of the open 
set $dS^{\pm}$. What kind of spaces lie in $\partial (dS^{\pm}) = 
\overline{dS^{\pm}} \setminus dS^{\pm}$, where the closure is 
taken in the smooth topology induced via (3.1)? Recall that solutions in 
$dS^{\pm}$ are conformally compact, and so conformally equivalent to a 
smooth bounded metric on the product $I\times \Sigma$. Generically, such 
metrics are "tall" in the sense of [36], [37], in that they contain 
Cauchy surfaces entirely visible to observers at sufficiently late 
times. A spacetime $(M, g) \in  dS^{+}$ (or $dS^{-}$) is called 
``infinitely tall'' if it is conformally equivalent to a complete, 
(but not necessarily product), metric on $[0,\infty)\times \Sigma$ 
with $\{0\}\times \Sigma  = {\mathcal I}^{+}$, (or ${\mathcal I}^{-}$).

\begin{theorem} \label{t 3.2.}
  An element in the boundary $\partial dS^{\pm}$ of $dS^{\pm}$, for 
$n = 3$, is given by exactly one of the following three configurations:

 I. A pair of solutions $(M, g^{+}) \in  dS^{+}$ and $(M, g^{-}) \in  
dS^{-}$, each geodesically complete and globally hyperbolic. One has 
${\mathcal I}^{-} = \emptyset$ for $(M, g^{+})$ and ${\mathcal I}^{+} = 
\emptyset$ for $(M, g^{-})$. Both solutions $(M, g^{+})$ and $(M, 
g^{-})$ are infinitely tall, and "infinitely far apart". 

 II. A single geodesically complete and globally hyperbolic solution 
$(M, g) \in  dS^{+}$, either with a partial compactification at 
${\mathcal I}^{-}$, or ${\mathcal I}^{-} = \emptyset$.

 III. A single geodesically complete and globally hyperbolic solution 
$(M, g) \in  dS^{-}$, either with a partial compactification at 
${\mathcal I}^{+},$ or ${\mathcal I}^{+} = \emptyset$. 
\end{theorem}
\noindent
{\bf Proof:}
 One can work at either ${\mathcal I}^{+}$ or ${\mathcal I}^{-},$ and so we 
choose ${\mathcal I}^{+}.$ Given data $(g_{(0)}^{+,i}, g_{(3)}^{+,i}) \in  
{\mathcal C}^{+}$ giving solutions $(M, g^{i})$ in $dS^{\pm}$, suppose 
$g_{(0)}^{+,i} \rightarrow  g_{(0)}^{+},$ and $g_{(3)}^{+,i} 
\rightarrow  g_{(3)}^{+}$ in the $C^{k}$, ($k \geq 7$ for instance), 
topology on ${\mathcal C}^{+}$, (or analytic topology for analytic data). 
The induced data on ${\mathcal I}^{-}$ may either converge to a limit, 
(in a subsequence), or diverge to infinity in ${\mathcal C}^{-}$.

 Suppose first the data on ${\mathcal I}^{-}$ converge to a limit, so that
\begin{equation} \label{e3.4}
g_{(0)}^{-,i} \rightarrow  g_{(0)}^{-}, \ \ {\rm and} \ \  g_{(3)}^{-,i} 
\rightarrow  g_{(3)}^{-}. 
\end{equation}
Of course, by definition, the metrics $(M, g^{i})$ do not converge to a 
limit $(M, g) \in  dS^{\pm}$.

 By Friedrich's theorem [12], there exist maximal globally hyperbolic 
$dS^{+}$ and $dS^{-}$ solutions $(M, g^{+})$ and $(M, g^{-})$, defined 
at least for $(T, \infty)$ and $(-\infty , -T)$ respectively, which realize 
the limit boundary data at ${\mathcal I}^{+}$ and ${\mathcal I}^{-}$. Here $T = 
-\log \rho_{0}$, where $\rho^{\pm}$ is the geodesic defining function for 
${\mathcal I}^{+}$ or ${\mathcal I}^{-}$, and $\rho^{\pm} = \rho_{0}$ is small. 
Further, on $(\rho^{+})^{-1}(T, \infty)$ and $(\rho^{-})^{-1}(-\infty , -T)$, 
the metrics $g^{i} \rightarrow  g^{+}$ and $g^{i} \rightarrow  g^{-}$ 
respectively. (This fact is the only place in the proof where the 
condition $n = 3$ is needed).

 Let $S_{i}(T^{+})$ be the $T^{+}$ level set of $t^{+} = -\log\rho^{+}$ 
with respect to $g^{i}$ and similarly $S_{i}(T^{-})$ the $T^{-}$ level 
set of $t^{-} = \log \rho^{-}$. We claim that the geodesic distance 
between $S_{i}(T^{-})$ and $S_{i}(T^{+})$ diverges to $\infty$, as 
$i \rightarrow  \infty$, i.e. all timelike maximal geodesics joining 
$S_{i}(T^{-})$ to $S_{i}(T^{+})$ have length diverging to $\infty$ as 
$i \rightarrow \infty$. To see this, the Cauchy data of $g^{i}$ at 
$S_{i}(T^{-})$ and $S_{i}(T^{+})$ converge to the Cauchy data of $g^{-}$ 
at $S(T^{-})$ and the Cauchy data of $g^{+}$ at $S(T^{+})$ respectively. 
Since the solutions $g^{i}$ exist from time $T^{-}$ to $T^{+}$, 
(and much further), the Cauchy stability theorem for the vacuum Einstein 
equations implies that the limit solution $g$ exists between $S(T^{-})$ and 
$S(T^{+})$ if the distance between $S_{i}(T^{-})$ and $S_{i}(T^{+})$ remains 
uniformly bounded, (at all points). Hence, $g^{+}$ and $g^{-}$ are part of a 
global solution $(M, g) \in  dS^{\pm}$, giving a contradiction. It follows 
that the limit solutions $g^{+}$ and $g^{-}$ are infinitely far apart.

 Exactly the same reasoning, choosing suitable $T^{-}$ or $T^{+}$, proves that 
each maximal globally hyperbolic spacetime $(M, g^{+})$ and $(M, g^{-})$ 
is geodesically complete. 

 Next, we need to show that $g^{+}$ has ${\mathcal I}^{-} = \emptyset  .$ 
(The argument that ${\mathcal I}^{+} = \emptyset  $ for $g^{-}$ is the 
same). This will also prove that $g^{+}$ is infinitely tall. Suppose 
instead ${\mathcal I}^{-} \neq  \emptyset  ,$ so that there is a least a 
partial compactification ${\mathcal I}^{-}$ of $g^{+}$ at past infinity. 

 For $i$ sufficiently large, the metrics $g^{i}$ are (arbitrarily) 
close to the limit $g^{+}$ on (arbitrarily) large regions of $g^{+}.$ 
Of course, by the analysis above, the metrics $g^{i}$ extend "much 
further" to the past of $g^{+}$. Fix $i$ large for the moment, and let 
$\rho  = \rho_{i}^{-}$ be the geodesic defining function for past 
infinity $({\mathcal I}^{-}, g_{(0),}^{-,i})$ of $(M, g^{i})$. Let $H$ be 
the mean curvature of the level sets $S(\rho)$ of $\rho$ with respect to 
$g^{i}$. By [34], the ratio
\begin{equation} \label{e3.5}
\frac{3-H}{\rho^{2}} = \frac{\bar H}{\rho} 
\end{equation}
is monotone decreasing in $\rho$; in fact if $\phi = \frac{\bar H}{\rho}$, 
then 
\begin{equation} \label{e3.6}
\phi' \leq -\frac{\rho \phi^{2}}{3} \leq 0 .
\end{equation} 
Here $\bar H$ is the mean curvature of the level sets $S(\rho)$ with respect 
to the compactified metric $\bar g = \bar g^{i}$ as in (2.1); $H$ is taken with 
respect to the future, while $\bar H$ is taken with respect to the past unit normal, 
so $H > 0$ corresponds to expansion to the future. 

  Note that the differential inequality (3.6) implies that if $\phi (\rho_{0}) 
< 0$ for some $\rho_{0}$, then $\phi \rightarrow -\infty$ at $\rho_{0} + r$, 
for some $r \leq r_{0}$, where $r_{0}$ depends only on $\phi (\rho_{0})$. Thus, 
at least on those $\rho$-geodesics which on $(M, g^{i})$ which extend, (as maximal 
curves in the $\rho$-congruence), all the way to ${\mathcal I}^{+}$, the inequality 
(3.6) implies $\phi(\rho) > 0$ for all $\rho$. It then also follows that $\phi(\rho)$ 
becomes arbitrarily small, for $\rho$ sufficiently large. Note that the set of such 
geodesics contains a non-empty open set ${\mathcal V}$ in $(M, g^{i})$; of course 
${\mathcal V}$ changes with the choice of defining function $\rho$, i.e. the choice 
of boundary metric on ${\mathcal I}^{-}$ for $(M, g^{i})$. 

 Now choose $T = T_{i}$ large enough so that $S(T) \subset (M, g_{i})$ 
is partially close to a domain in $g^{+}$ near past infinity 
${\mathcal I}^{-}$ in $g^{+}$. Thus there is a domain $U \subset S(T) 
\subset (M, g^{i})$ such that $(U, \bar g^{+})$ is close to a 
domain $U_{\infty} \subset ({\mathcal I}^{-}, g_{(0)}^{-})$. Since $\rho$ 
is extremely large on $S(T)$, the ratio in (3.5) is very small on the open 
subset $S(T)\cap {\mathcal V}$. However, (3.5) holds for any geodesic 
compactification and one can renormalize $\rho$ by sending $\rho \rightarrow  
\varepsilon\rho  = \widetilde \rho$. This change of compactification makes the 
metric $g_{(0)}^{-,i}$ on ${\mathcal I}^{-}$ very small; $\widetilde g_{(0)}^{-,i} = 
\varepsilon^{2}g_{(0)}^{-,i}$. Setting $\varepsilon  = T^{-1}$ then gives 
$\widetilde \rho = 1$ on $S(T)$. The numerator $3-H$ in (3.5) is of course 
``scale-invariant''; it does not depend on scale of $\rho$, since it depends 
only on $(M, g_{i})$, and so does not involve any compactified metric. 

 It follows now from the choice of $T$ that the ratio
\begin{equation} \label{e3.7}
\frac{3-H}{\widetilde \rho^{2}} 
\end{equation}
in (3.5) is very small; it can be made arbitrarily small by choosing 
$i$ sufficiently large and $T$ sufficiently large so that $U \subset 
S(T)$ is sufficiently close to $U_{\infty}$. Now the monotonicity of 
(3.5), (applied on each $g^{i}$), implies that as $\widetilde \rho$ 
increases, the ratio (3.7) becomes even smaller. 

 It follows that in the limit, the domain in $g^{+}$ formed by 
$\rho$-curves between ${\mathcal I}^{-}$ and ${\mathcal I}^{+}$ has 
$H \equiv n = 3$. Via standard arguments with the Raychaudhuri equation, 
it follows that this region is isometric to pure de Sitter space, or a 
finite quotient thereof. Since $g^{+}$ is globally hyperbolic and geodesically 
complete, it then follows by similar arguments that $g^{+}$ is globally isometric 
to pure de Sitter space, or a finite quotient thereof. Hence, the limit data 
$(g_{(0)}^{+}, g_{(3)}^{+})$ equals the data $(\gamma_{0}, 0)$ of pure de Sitter 
space, (or a quotient), a contradiction.

\medskip

 Next suppose (3.4) fails, so that either $g_{(0)}^{i,-}$ or 
$g_{(3)}^{i,-}$ diverges to $\infty$ on ${\mathcal I}^{-}$. The same 
reasoning as before shows that the limit $(M, g) \in  dS^{+}$ is 
geodesically complete and globally hyperbolic. In this setting, one 
then has only a partial past infinity ${\mathcal I}^{-}$, given by the 
domain (possibly empty) on which both $g_{(0)}^{i,-}$ and 
$g_{(3)}^{i,-}$ converge to the limit $(g_{(0)}^{-}, g_{(3)}^{-}).$ 
This situation then gives Case II. Case III is obtained in the same 
way, by interchanging ${\mathcal I}^{-}$ with ${\mathcal I}^{+}.$
{\endproof}

 Exactly the same proof holds in all dimensions if an openness or 
Cauchy stability result holds, as in the case $n = 3$.

\medskip

 Examples discussed in \S 4, (the Taub-NUT curve) show that the 
configuration in Case I does occur. On the other hand, there are no 
known examples where the configuration in Case II or III above occurs, 
and it would be interesting to know if such configurations can actually 
arise or not. 

 It would also be interesting to prove that the completion 
${\overline{dS^{\pm}}}$ is compact in the topology on ${\mathcal A}^{+}$ for 
instance, i.e. if the final data $(\gamma^{+}, g_{(n)}^{+})$ on ${\mathcal 
I}^{+}$ are sufficiently large, then the corresponding solution cannot 
be in $dS^{\pm}.$ The condition that $(M, g) \in dS^{\pm}$ implies 
$R_{\gamma^{+}} > 0$, (and $R_{\gamma^{-}} > 0$), is one measure of 
this. Finally, one would like to know if the space $dS^{\pm}$ is 
connected, as a domain in ${\mathcal A}^{+}$ for instance.

 These questions correspond to the picture that $dS^{\pm}$ is given by 
a connected, bounded domain in ${\mathcal A}^{+}$, (or ${\mathcal A}^{-}$), with 
a wall or boundary described by Theorem 3.1. For initial data on 
${\mathcal A}^{-}$ for instance, one expects the formation of black holes 
and big bang type singularities outside the wall. 

 Finally, consider limits $(M, g) \in  dS^{+}$ of $dS^{\pm}$ which have 
${\mathcal I}^{-} = \emptyset$. For simplicity, assume $n = 3$, so that, by 
[33] and [35] as mentioned above, $M$ is then topologically of the form 
${\mathbb R}\times S^{3}$, (in a finite cover). Thus, as $T \rightarrow  
-\infty$, the Cauchy surfaces $S(T)$ degenerate, when compactified. The 
simplest example of a diverging curve metrics on $S^{3}$ is a collapsing 
family, where the lengths of the circle fibers $S^{1}$ in the Hopf fibration 
$S^{3} \rightarrow  S^{2}$ have length converging to 0. This type of 
degeneration occurs with uniformly bounded curvature tensor, and is 
illustrated concretely on the Taub-NUT curve, discussed below.

\section{Examples.}
\setcounter{equation}{0}

 In this section, we discuss several classes of examples illlustrating 
the work in \S 3.

{\bf Example 4.1.}
 Let $g_{TN}$ be the curve of AH (or EAdS) Taub-NUT metrics on the 4-ball 
$B^{4},$ cf. [38] for instance, given by
\begin{equation} \label{e4.1}
g_{TN} = l^{2}E[\frac{(r^{2}-1)}{F(r)}dr^{2} + \frac{F(r)}{(r^{2}-1)}\theta_{1}^{2} + 
\frac{(r^{2}-1)}{4}g_{S^{2}(1)}], 
\end{equation}
where $E \in  (0,\infty )$ is any constant, $r \geq $ 1, and 
\begin{equation} \label{e4.2}
F(r) = Er^{4} + (4-6E)r^{2} + (8E-8)r + 4-3E = 
(r-1)^{2}\{E(r+1)^{2}+4(1-E)\} > 0. 
\end{equation}
Here $\theta_{1} \in  [0,2\pi]$ parametrizes the circle $S^{1}$ in the 
Hopf fibration $S^{3} \rightarrow  S^{2}$. The nut charge description of 
$g_{TN},$ cf. [39] for instance, given by
\begin{equation} \label{e4.3}
g_{TN} = V^{-1}dr^{2} + V\theta_{1}^{2} + (r^{2}-n^{2})g_{S^{2}}, 
\end{equation}
where
$$V(r) = \frac{(r^{2}+n^{2}) - 2mr + 
l^{-2}(r^{4}-6n^{2}r^{2}-3n^{4})}{(r^{2}-n^{2})}, $$
is equivalent to (4.1) under the substitution $r \rightarrow nr$, with 
$n^{2} = \frac{l^{2}E}{4}$, with mass parameter $m$ given by
\begin{equation} \label{e4.4}
m = \frac{l}{2}E^{1/2}(1-E). 
\end{equation}
The AH Taub-NUT metric is self-dual Einstein and has conformal infinity $\gamma$ 
given by the ``Berger sphere'' with $S^{1}$ fibers of length $\beta  = 2\pi 
E^{1/2}\in  (0,\infty )$ over $S^{2}(\frac{1}{2})$. The scalar curvature 
$R_{\gamma}$ of the boundary metric $\gamma $ satisfies
$$R = 8 - 2E.$$
Clearly $\gamma$ is real-analytic, as is the geodesic compactification with 
boundary metric $\gamma$. Since $g$ is self-dual, [40] implies that the 
$g_{(3)}$ term in the expansion (2.6), (see also (5.3) below), is given by 
\begin{equation} \label{e4.5}
g_{(3)} = \tfrac{1}{3}*dRic, 
\end{equation}
wher $Ric = Ric_{\gamma}$ is viewed as a vector valued 1-form, $d$ is the exterior 
derivative and $*$ the Hodge $*$ operator, all with respect to $\gamma$. When 
$E = 1$, $g_{TN}$ is the Poincar\'e metric.

\medskip

 The de Sitter continuation of $g_{TN}$ at ${\mathcal I}^{+}$ is the dS 
Taub-NUT metric on ${\mathbb R}\times S^{3}$, cf. [41] for instance, given by
\begin{equation} \label{e4.6}
g^{TN} = l^{2}E[-\frac{(\tau^{2}+1)}{A(\tau )}d\tau^{2} + \frac{A(\tau)}
{(\tau^{2}+1)}\theta_{1}^{2} + \frac{(\tau^{2}+1)}{4}g_{S^{2}}], 
\end{equation}
where $\tau\in (-\infty ,\infty)$ and
\begin{equation} \label{e4.7}
A(\tau) = E\tau^{4} - (4-6E)\tau^{2} - (8E-8)\tau  + 4-3E. 
\end{equation}
Again when $E = 1$, $g^{dS}$ is the (exact) de Sitter metric. Changing 
$n \rightarrow in$, $r \rightarrow  i\tau$ transforms (4.3) into the 
nut form of $g^{TN}$.

 The dS Taub-NUT metric $g^{TN}$ is complete and globally hyperbolic, 
without singularities, exactly when $A(\tau) > 0$, for all $\tau$. A 
straightforward but lengthy calculation shows this is the case if and 
only if
\begin{equation} \label{e4.8}
E \in  [\frac{2}{3}, \frac{1}{3}(2+ \sqrt{3})]. 
\end{equation}

  For $E$ in the range (4.8), there are no closed time-like curves, in 
contrast to the AdS (or $\Lambda = 0$) Taub-NUT metric, which always 
have such curves. (Friedrich's theorem [12] also guarantees the existence 
of some interval $|E - 1| < \varepsilon$ which has the same overall global 
structure of pure de Sitter space).

 Consider first the situation $E \in  (\frac{2}{3}, \frac{1}{3}(2+\sqrt{3}))$. 
Then both ${\mathcal I}^{+}$ and ${\mathcal I}^{-}$ are well-defined and $g^{TN} = 
g^{TN}(E) \in dS^{\pm}$. Observe from the explicit form of (4.6) that
\begin{equation} \label{e4.9}
\gamma^{-} = \gamma^{+} \ \ {\rm and} \ \  g_{(3)}^{+} = -g_{(3)}^{-}; 
\end{equation}
here $g_{(3)}^{+}$ is taken with respect to the past timelike direction 
$\partial_{\rho},$ while $g_{(3)}^{-}$ is taken with respect to the 
future timelike direction $\partial_{\rho},$ Thus, even though $g^{TN}$ 
is not time-symmetric when $E \neq $ 1, there is no gravitational 
scattering from past to future conformal infinity; from (3.3), one has
$${\mathcal S}  = id, $$
on the Taub-NUT curve. The AH continuation of the dS Taub-NUT metric at 
${\mathcal I}^{-}$ is the anti self-dual AH Taub-NUT metric (4.3), (due to 
the change in orientation), with $n$ replaced by $-n$, changing $m$ to $-m$; 
this metric has an isolated cone-like singularity at the origin $\{r = n\}$. 

\medskip

 Next consider the extreme values $E_{-} = \frac{2}{3}$ and $E_{+} = 
\frac{1}{3}(2+\sqrt{3})$, and let $\tau_{-} = -1$, $\tau_{+} = (2-\sqrt{3})$. 
At $E = E_{-}$, $A(\tau) \geq 0$ for all $\tau$, with $A(\tau) = 0$ only at 
$\tau_{-}$. Moreover, $\tau_{-}$ is a degenerate zero in that 
$A'(\tau_{-}) = 0$. Exactly the same remarks hold at $E_{+}$ and 
$\tau_{+}$. At the extreme values $E_{\pm}$, the metric $g^{TN} = 
g^{TN}(E_{\pm})$ is still geodesically complete and globally 
hyperbolic. However, neither of these two metrics is in $dS^{\pm}$. At 
$E_{-}$ for instance, there are two solutions $g^{TN}_{p}$ and 
$g^{TN}_{f}$, parametrized by $(-\infty , \tau_{-})$ and $(\tau_{-}, 
\infty)$ respectively, each complete and globally hyperbolic, with 
$g^{TN}_{p} \in  dS^{-}$, $g^{TN}_{f} \in  dS^{+}$, with same conformal 
infinities $\gamma^{+} = \gamma^{-} = \gamma (E_{-})$ on each end. This 
situation corresponds of course to Case I in Theorem 3.1.

 As one goes to future infinity in $g^{TN}_{p}$, or past infinity in 
$g^{TN}_{f}$, the spatial metric degenerates, by collapsing the $S^{1}$ 
fiber in the Hopf fibration to a point, while the radius of the base 
$S^{2}$ converges to $(E_{-}(\tau_{-}^{2}+1)/4)^{1/2}$. The curvature of 
these metrics is uniformly bounded. Exactly the same structure holds at 
$E_{+}$. 

 The two extreme metrics $g^{TN}_{p}$ and $g^{TN}_{f}$ are analogous to 
extreme black hole solutions. For $E$ outside the range (4.8), $A(\tau)$ 
acquires up to 4 zeros, so as $\tau$ ranges over all ${\mathbb R}$, 
one has several copies of the Taub-NUT metrics glued along horizons 
by analytic continuation; cf. [19] for a detailed description.

\medskip

 One expects that exactly the same behavior holds for all dS Bianchi IX 
metrics, and it would be interesting to see if this is the case. The 
corresponding AdS Bianchi IX metrics are also self-dual, and have been 
described in detail in [30]. 

 There are Taub-NUT metrics in higher (even) dimensions, cf. [19], [42], 
and the discussion above holds in essentially the same form.

  As a final remark, as $E \rightarrow \infty$, the AH Taub-NUT metrics 
limit on the Bergmann or complex hyperbolic metric on the unit ball in 
${\mathbb C}^{2}$; the corresponding conformal infinity is degenerate, cf. 
[43] for example. This limit corresponds to taking $R_{\gamma} \rightarrow 
-\infty$, and hence by (3.2), there is no dS continuation of the 
Bergmann metric. 

\medskip
\noindent
{\bf Examples 4.2.}
 The AH Schwarzschild metrics are a curve of metrics on $M = {\mathbb R}^{2}
\times S^{n-1}$ given by
\begin{equation} \label{e4.10}
g_{Sch} = V^{-1}dr^{2} + Vd\theta^{2} + 
(\frac{r}{l})^{2}g_{S^{n-1}(l)}, 
\end{equation}
where 
\begin{equation} \label{e4.11}
V(r) = 1+(\frac{r}{l})^{2}-\frac{\mu}{r^{n-2}}, 
\end{equation}
Here $r \in  [r_{+},\infty)$, where $r_{+}$ is the largest root of $V$, 
and the circular parameter $\theta \in [0,\beta]$, 
$\beta  = 4\pi l^{2}r_{+}/(nr_{+}^{2}+(n-2)l^{2})$, with $\mu  = 2mG$. 
It is easy to see that the conformal infinity of $g_{Sch}(m)$ is given by the 
conformal class of the product metric $\gamma $ on $S^{1}(\beta)
\times S^{n-1}(l)$. As a function of $m \in (0,\infty)$, observe that 
$\beta$ has a maximum value of $\beta_{0} = 4\pi l^{2}/(1+l^{2})(n(n-2))^{1/2}$, 
and for every $m \neq  m_{0}$, there are two values $m^{\pm}$ of $m$ giving 
the same value of $\beta$. Thus two distinct metrics have the same conformal 
infinity.

 The de Sitter continuation of the AH Schwarzschild metric is not 
exactly the dS Schwarzschild metric. Suppose first $n = 3$. Then the 
continuation (at ${\mathcal I}^{+})$ is the dS Kantowski-Sachs metric on 
${\mathbb R}\times S^{1}\times S^{2}$, given by 
\begin{equation} \label{e4.12}
g^{KS} = -V^{-1}dt^{2} + \alpha^{2}V d\theta^{2} + t^{2}g_{S^{2}(l)}, 
\end{equation}
where 
\begin{equation} \label{e4.13}
V = V(t) = t^{2} - 1 + \frac{\mu}{t}, 
\end{equation}
and we have set $l = 1$. Here $\alpha > 0$ is a free parameter; a continuous 
continuation from $g_{Sch}$ to $g^{KS}$ then requires $\alpha  = 1$ but $g^{KS}$ 
is defined for any $\alpha > 0$; see [44] for a general discussion of these 
metrics. 

 By the result of Andersson-Galloway [33] discussed in \S 3, 
${\mathcal I}^{-} = \emptyset$ for these metrics, i.e. for the maximal globally 
hyperbolic developments. To see this in detail, consider first the case 
$\mu  = 0$; note however that the AH Schwarzschild metric $g_{Sch}$ with 
$\mu  = 0$ is degenerate, i.e. not defined. The metric $g^{KS}(0)$ is a 
quotient of a domain in pure de Sitter space by a discrete ${\mathbb Z}$-action. 
The Penrose diagram for pure de Sitter space is a square with the $t$-level 
curves in (4.12) given by hyperbolas in the upper triangle when $t > 1$; for 
$t = 1$, the level curves are the diagonals of the square. The ${\mathbb Z}$-action 
is a boost symmetry (of length determined by $\alpha \cdot \beta$). Near the horizon 
$t = 1$, the geometry is that of a round 2-sphere $t^{2}g_{S^{2}(1)}$ times the Misner 
spacetime, cf. [8]. Thus, there are analytic continuations across the 
horizon, giving rise to closed time-like curves. When $\mu > 0$, these closed 
time-like curves persist, but there is now in addition a singularity at $t = 0$. 
In both cases, the maximal globally hyperbolic spacetime has ${\mathcal I}^{-} = 
\emptyset$. 

\medskip

 The universal cover of the dS Kantowski-Sachs solutions are the dS Schwarzschild 
metrics; in terms of (4.12), one just replaces the circular parameter 
$\theta  \in  [0,\beta]$ by $r \in  {\mathbb R}$; then (4.12) becomes the 
usual form of the dS Schwarzschild metric in the exterior of the 
cosmological horizon. When $\mu = 0$, this gives the pure de Sitter metric 
outside the horizon, while the extension inside the horizon is de Sitter 
in static coordinates, given by, (interchanging $t$ and $r$ coordinates), 
\begin{equation} \label{e4.14}
g_{dS} = -(1-r^{2})dt^{2} + (1-r^{2})^{-1}dr^{2} + r^{2}g_{S^{2}(1)}. 
\end{equation}

 Future conformal infinity ${\mathcal I}^{+}$ has thus changed from the 
compact manifold $S^{1}\times S^{2}$ to the universal cover ${\mathbb R}\times 
S^{2}$. Of course, this can be conformally changed to $S^{3} \setminus 
\{q\cup -q\}$, where $q$ and $-q$ are two antipodal points, and so extended 
to $S^{3}$, giving rise to the usual compactification of pure de Sitter in 
global coordinates. The translation Killing field, (the analytic continuation 
of the static time-like Killing field $\partial_{t}$ in (4.14)), then becomes 
a conformal Killing field, (the dilatation field) on $S^{3}$.

 When $\mu > 0$, one again has ${\mathcal I}^{+} = {\mathbb R}\times S^{2}$. 
However, the singularity at $t = 0$ in (4.12) now propagates to future infinity, 
giving a 2-sphere of 0 radius times  ${\mathbb R}$ at ${\mathcal I}^{+}$. 
Analytic continuation through horizons gives the usual infinite Penrose 
diagram for the maximally extended dS Schwarzschild metrics, cf. [1]. 

\medskip

 On the AH side, ${\mathbb R}^{2}\times S^{2}$ is simply connected, so the 
$S^{1}$ at infinity cannot be unwrapped. Consider however the AdS 
Schwarzschild metric obtained from the AH Schwarzschild metric (4.10) 
by replacing $\theta$ by $it$;
\begin{equation} \label{e4.15}
g_{Sch} = -Vdt^{2} + V^{-1}dr^{2} + r^{2}g_{S^{2}(1)}. 
\end{equation}
This static metric has conformal infinity of the form ${\mathcal I}^{+} = 
{\mathbb R}\times S^{2}$, so its continuation across ${\mathcal I}^{+}$ is the dS 
Schwarzschild metric. 

\medskip

 Now consider the same situation with $n > 3$. The continuation of an 
AH Schwarzschild metric is again a Kantowski-Sachs metric of the form 
(4.12), (with $S^{n-1}(1)$ in place of $S^{2}(1)$), and with $V$ of the form
\begin{equation} \label{e4.16}
V = V(t) = t^{2} - 1 + \frac{\mu}{t^{n-2}}. 
\end{equation}
The terms $\mu /r^{n-2}$ in (4.11) and $\mu /t^{n-2}$ in (4.16) essentially 
determine the $g_{(n)}$ term in the Fefferman-Graham-Starobinsky 
expansion, or more precisely its deviation from the $g_{(n)}$ term for 
pure AH or dS space. The relation (2.17) of the $g_{(n)}$ terms in the AH 
and dS expansions then requires that
$$\mu_{dS} = \pm\mu_{AH}, $$
where $+$ occurs if $n \equiv 2, 3$ (mod 4) and $-$ occurs if $n \equiv 
0, 1$ (mod 4). Thus, the masses $\mu$ agree, and are both positive, 
only when $n \equiv $ 2, 3 (mod 4). When $n = 4$ for example, the 
positive mass AH Schwarzschild metric continues to the negative mass 
Kantowski-Sacks metric and vice versa. These negative mass metrics of 
course have (naked) curvature singularities; cf. \S 5 for further 
discussion.

\medskip

 The discussion above generalizes easily to higher genus black holes, 
by replacing the constant 1 by $k$, $k = 0$ or $-1$, and $S^{2}$ with the 
torus $T^{2}$ or a surface of higher genus $\Sigma_{g}$ respectively. 
Moreover, these spaces may be replaced by Einstein metrics on any 
$(n-1)$-dimensional manifold, cf. [39], [45].

\section{Holographic Renormalization.}
\setcounter{equation}{0}

 Einstein metrics (2.2) are critical points of the Einstein-Hilbert 
action. For Riemannian (i.e Euclidean) metrics, the action is usually 
taken to be
\begin{equation} \label{e5.1}
I = -\frac{1}{16\pi G}\int_{M}(R-2\Lambda )dv - \frac{1}{8\pi 
G}\int_{\partial M}H dA,  
\end{equation}
where $H$ is the mean curvature of the boundary. (The sign conventions 
are based on path integral considerations, cf. [46]). However, both 
terms in (5.1) are infinite on AH Einstein metrics. The main idea in 
the method of holographic renormalization is that one may find natural 
counterterms, depending only on the intrinsic geometry of the boundary 
metric $\gamma$, (or the metric on the cutoff at $t = \varepsilon$), such 
that if $I$ is renormalized by subtraction of these counterterms, the 
renormalized action $I^{ren}$ is finite, cf. [13], [14]. (Very briefly, the 
counterterms are obtained from the expansion of the action determined by 
(2.6) or (2.8)).  If $n$ is odd, $I^{ren}$ depends only on $(M, g)$, and 
not on any particular choice of compactification. However, if $n$ is even, 
$I^{ren}$ does depend on the compactification, i.e. on boundary metric 
$\gamma$, and not just on the conformal class $[\gamma]$. This behavior 
is closely related to the absence or presence of the conformal anomaly 
in odd and even dimensions respectively.

 The variation of $I^{ren}$ at a given AH Einstein metric $g$ is given 
by
\begin{equation} \label{e5.2}
dI^{ren}(h) = \frac{d}{dt}I^{ren}(g+th), 
\end{equation}
where $h$ is tangent to the space $E$ of AH Einstein metrics; (this is 
a smooth manifold, cf. [25]). The differential $dI^{ren}$ is the 
holographic stress-energy tensor; it is a 1-form on $E$  and 
corresponds to the renormalization of the quasi-local Brown-York stress-energy 
[47]. Via the AdS/CFT correspondence, $T = 2dI^{ren}$ gives the 
expectation value of the stress-energy tensor of the CFT on $\partial M$, 
in all known cases. 

 Since Einstein metrics are critical points of $I$ or $I^{ren},$ it is 
clear that $dI^{ren}$ must be supported on $\partial M.$ If $n$ is odd, 
it is proved in [14], cf. also [40], that
\begin{equation} \label{e5.3}
T = 2dI^{ren} = -\frac{n}{16\pi G}g_{(n)}, 
\end{equation}
while if $n$ is even,
\begin{equation} \label{e5.4}
T = 2dI^{ren} = -\frac{n}{16\pi G}(g_{(n)} + r_{(n)}), 
\end{equation}
where $r_{(n)}$ is explicitly determined by $\gamma$ and covariant 
derivatives of its curvature; it depends only on the lower order terms 
in the FG expansion. Thus
\begin{equation} \label{e5.5}
\langle T, h \rangle = -\frac{n}{16\pi G}\int_{\partial M}\langle g_{(n)} + 
r_{(n)}, h_{(0)} \rangle dv_{\gamma}, 
\end{equation}
where $h_{(0)}$ is the variation of the boundary metric $\gamma$ 
induced by $h$. The complexity of the term $r_{(n)}$ grows rapidly in $n$; 
explicit expressions for $n = 2, 4, 6$ are given in [14]. (The sign in 
(5.3) or (5.4) differs from that in [14], but agrees with the signs in 
[13], [39], [40]. With this choice of sign, the mass of pure AH$_{5}$ = 
EAdS$_{5}$ is positive in static coordinates $(\partial M = 
S^{1}\times S^{3})$).

 For Lorentzian metrics, the action has the form
\begin{equation} \label{e5.6}
I = \frac{1}{16\pi G}\int_{M}(R-2\Lambda )dv + \frac{1}{8\pi 
G}\int_{\partial M}H dA.  
\end{equation}
Note that this has the opposite sign to (5.1). For general AdS metrics, 
this action cannot be renormalized to a finite expression as in the AH 
case, since such metrics are not conformally compact, and in general 
are time dependent, with $t\in{\mathbb R} .$ The action can be 
renormalized to a finite value however for stationary AdS metrics.

 In any case, for general AdS metrics one does have a renormalized 
stress-energy tensor $T_{\mu\nu}.$ This is again given by (5.3), (5.4) 
but with the opposite sign. The stress-energy tensors $T^{AdS}$ of a 
static AdS metric $(M, g)$, $M = {\mathbb R}\times \Sigma$, and the 
corresponding $T^{AH}$ of the AH metric $(M, g)$ obtained by setting 
$\theta  = it$ are then related by
\begin{equation} \label{e5.7}
T^{AdS}_{tt} = T^{AH}_{\theta\theta}, \ \ T^{AdS}_{\Sigma} = 
-T^{AH}_{\Sigma}. 
\end{equation}

 To illustrate on a concrete example, let $g_{Sch}^{AH}$ and 
$g_{Sch}^{AdS}$ be the AH and AdS Schwarzschild metrics respectively, 
with static compactification. Then the definitions above give
\begin{equation} \label{e5.8}
T_{Sch}^{AH} = -\frac{1}{16\pi G}(\frac{\mu}{l^{n-2}} + 
\frac{2c_{n}}{n-1})((1-n)d\theta^{2} + g_{S^{n-1}(1)}), 
\end{equation}
while 
\begin{equation} \label{e5.9}
T_{Sch}^{AdS} = \frac{1}{16\pi G}(\frac{\mu}{l^{n-2}} + 
\frac{2c_{n}}{n-1})(-(1-n)dt^{2} + g_{S^{n-1}(1)}), 
\end{equation}
where $c_{n} =$ 0 if $n$ is odd and $c_{n} = 
(-1)^{n/2}\frac{(n-1)!!^{2}}{n!}$ if $n$ is even, cf. [39] and [45].

 On the de Sitter side, since spatial infinity is compact, the action 
(5.6) can be holographically renormalized in exactly the same way as on 
the AH side, and the same renormalization procedure gives a holographic 
stress-energy tensor $T^{dS} = 2dI^{ren}$. Because the AH action (5.1) and 
dS action (5.6) are both positive, one thus has
\begin{equation} \label{e5.10}
T^{dS}(g^{dS}) = T^{AH}(g^{AH}), 
\end{equation}
and $T^{dS}$ is given by (5.3) or (5.4), with the de Sitter terms 
$g_{(k)}^{dS}$ in place of the AH terms $g_{(k)}^{AH}.$ 

 If now $g^{dS}$ is the analytic continuation of $g^{AH}$ in the sense 
of Theorem 2.1, the formula (2.17) implies 
\begin{equation} \label{e5.11}
T^{dS} = -T^{AH}, \ {\rm if} \  n \equiv  3(mod 4),  \ \ {\rm and} \ \ 
T^{dS} = T^{AH}, \ {\rm if} \  n \equiv  1(mod 4).  
\end{equation}
Similarly, 
\begin{equation} \label{e5.12}
T^{dS} = T^{AH}, \ {\rm if} \ n = 4,  \ \ {\rm and} \ \ 
T^{dS} = -T^{AH}, \ {\rm if} \  n = 2.  
\end{equation}
Curiously, it is not clear if analogous formulas hold in general for $n$ even, 
$n \geq 6$, since the term $r_{(n)}$ does not seem to have a definite 
sign change under the correspondence (2.18) in this range, according to 
[14].

\medskip

 We now turn to a discussion of conserved quantities of AH and dS 
metrics associated with conformal Killing fields at conformal infinity, 
following [13]-[15]. First, it should be noted that a generic 
conformal infinity $(\partial M, [\gamma])$ has no conformal Killing fields, 
and so the discussion applies at the outset only to a restricted class of spaces.

 We begin in the AH setting. Let $(M, g)$ be an AH Einstein metric with 
conformal infinity $(\partial M, \gamma ).$ One may define a conserved 
quantity associated with any conformal Killing field $K$ on $(\partial 
M, \gamma )$ by
\begin{equation} \label{e5.13}
Q = \int_{S}\langle T(K), \nu \rangle, 
\end{equation}
where $T$ is the holographic stress-energy tensor, $S$ is a slice to 
the orbits of the Killing field on $(\partial M, \gamma)$, and $\nu$ is 
the unit normal to $S$, in the direction $K$. The integral (5.13) is 
independent of $S$, since $T$ is transverse-traceless. If $n$ is odd, 
$Q$ depends only on the Einstein metric $(M, g)$ and choice of Killing 
field $K$ on $(\partial M, \gamma)$. However, if $n$ is even, $Q$ 
depends on the choice of boundary metric $\gamma  \in  [\gamma]$. 

 As mentioned in \S 2, any conformal Killing field at conformal infinity 
extends to a Killing field of any (globally smooth) AH Einstein bulk 
metric $(M, g)$. Furthermore, at least in many circumstances, if $K$ is 
static, i.e. hypersurface orthogonal on $\partial M$, then $K$ is also 
static on $(M, g)$, cf. [48]. 

 The same definition (5.13) holds for AdS metrics $(M, g)$. In 
particular, if $(M, g)$ is stationary in a neighborhood of infinity, then 
the (holographically renormalized) mass of $(M, g)$ is given by
\begin{equation} \label{e5.14}
m = \int_{S}\langle T(K), \nu \rangle , 
\end{equation}
where $S$ is a spacelike slice for the Lorentz metric $\gamma$ on 
$\partial M$ and $K$ is the timelike Killing field. Again, the value of $m$ 
depends on the choice of boundary metric $\gamma$ in $[\gamma]$ when $n$ is even.

 In particular, if the AdS metric $(M, g)$ is globally static 
(near $\partial M$), then $(M, g)$ can be Wick rotated to an AH Einstein metric, 
($t \rightarrow  it = \theta$). The AH mass and the AdS mass are then equal 
by (5.7). While in general stationary AdS metrics cannot be Wick rotated to 
stationary AH metrics, this can be done in many interesting concrete 
cases, such as the AdS Taub-NUT and Kerr metrics. In such a 
situation, the masses again agree by (5.7).

\medskip

 Conserved quantities can be defined in same way as (5.13) for metrics 
$(M, g)$ in $dS^{+}$ or $dS^{-}$ having Killing fields at ${\mathcal I}^{+}$ 
or ${\mathcal I}^{-}$. Since ${\mathcal I}$ is spacelike, it is not 
immediately clear which of these quantities should correspond to mass, 
and which to angular momentum-type quantities. A natural proposal due to 
Balasubramanian, de Boer and Minic [15], is that $Q$ in (5.13) gives the 
mass of $(M, g)$ at ${\mathcal I}^{+}$ when the Killing field is the 
``analytic continuation'' of a timelike vector field of $(M, g)$, which is 
asymptotically Killing and static near ${\mathcal I}^{+}$. In more detail, 
since the metric (locally) asympotically approaches the de Sitter geometry, 
one can write the metric near ${\mathcal I}^{+}$ in an approximately static chart, 
as in (4.14), and then continue the approximate timelike Killing field into the 
exterior region of the cosmological horizon to obtain an approximate 
spacelike Killing field. When this process leads to a Killing field on 
${\mathcal I}^{+}$, then the associated conserved quantity is defined to be 
the mass. One difficulty in general is that if the data $(g_{(0)}, g_{(n)})$ 
at ${\mathcal I}^{+}$ are not analytic, it is not clear how to actually carry out 
an analytic continuation; there may be no coordinates in which the metric is 
analytic.

 Consider first the case of pure de Sitter spacetime. The exterior of the static 
chart leads to $({\mathcal I}^{+}, \gamma)$ given by the round product 
metric on ${\mathbb R}\times S^{n-1}(1)$, with Killing field given by 
translation along the ${\mathbb R}$-direction. In general, if $K = \partial_{s}$ 
is a non-vanishing conformal Killing field on ${\mathcal I}^{+}$, then locally 
the topology of ${\mathcal I}^{+}$ is $I\times S$, with metric of the form
\begin{equation} \label{e5.15}
\gamma  = e^{\phi s}[N^{2}ds^{2} + h_{ij}(dx_{i}+n_{i}ds 
)(dx_{j}+n_{j}ds )], 
\end{equation}
where $h$ is a metric on $S$, $N$, $n_{i}$ and $h_{ij}$ are independent of 
$s$ and $\phi : S \rightarrow  {\mathbb R}$. 

 Suppose that $K$ is static, so that $n_{i} = 0$. Then ${\mathcal I}^{+}$ 
is topologically of the form ${\mathbb R}\times S$, and the representation 
(5.15) holds globally over all ${\mathbb R}$. The continuation of $(M, g) \in  
dS^{+}$ across ${\mathcal I}^{+}$ gives an AH Einstein metric, with static 
Killing field $K$, defined at least in a neighborhood of conformal 
infinity $\partial M = {\mathcal I}^{+}$. By Wick rotation, this is 
equivalent to a static AdS Einstein metric, for which one has the 
natural mass definition given by (5.14). Thus, $K$ on ${\mathcal I}^{+}$ is 
time-like with respect to the naturally associated AdS metric. This 
notion of mass agrees with the BBM mass [15] discussed above. 

 Suppose next the Killing field $K$ on ${\mathcal I}^{+}$ is stationary and 
not static. Then the metric $\gamma$ on ${\mathcal I}^{+}$ either has the 
global form (5.15) on ${\mathbb R}\times S$, or ${\mathcal I}^{+}$ is a non-trivial 
$S^{1}$ bundle over $S$, with the fiber $S^{1}$ trivial, or of finite 
order, in $\pi_{1}({\mathcal I}^{+})$. In the latter case, one cannot 
unwrap the orbits of the Killing field to ${\mathbb R}$ by passing to a 
covering space. The AH Einstein continuation will also be stationary, 
and as mentioned above, in certain circumstances this can again be Wick 
rotated to a stationary AdS metric having the same mass. Note that in 
general the resulting AdS metric then has closed time-like curves at 
conformal infinity. For example, the dS Taub-NUT metric continues to the 
AH Taub-NUT metric which can be rotated to the AdS Taub-NUT metric. 

 The following result relates the mass of these two metrics.
\begin{proposition} \label{p 5.1.}
  Let $m_{dS}$ be the holographic mass of the de Sitter-type metric 
$(M, g)\in dS^{+},$ with static or stationary Killing field $K$ on 
${\mathcal I}^{+}$, and let $m_{AdS}$ be the mass of the associated AdS metric 
$(M, g)$. If $n$ is odd, then
\begin{equation} \label{e5.16}
m_{dS} = -m_{AdS}, \ {\rm if} \  n \equiv  3 (mod 4), \ \ {\rm and} \ \ 
m_{dS} = m_{AdS}, \ {\rm if}\  n \equiv  1 (mod 4).  
\end{equation}
Also,
\begin{equation} \label{e5.17}
m_{dS} = m_{AdS}, \ {\rm if} \  n = 2, \ \ {\rm and} \ \  
m_{dS} = -m_{AdS}, \ {\rm if} \ n = 4.  
\end{equation}

\end{proposition}
\noindent
{\bf Proof:}
 This follows immediately from the discussion above, together with 
(5.11)-(5.12).
{\endproof}

 We use these results to discuss the maximal mass conjecture proposed 
in [15], that if $(M, g) \in  dS^{+}$ has mass greater than the mass of 
pure de Sitter space, then $M$ has a cosmological singularity. A 
reasonable definition here is that $(M, g) \in  dS^{+}$ has a 
cosmological singularity if either $(M, g)$ contains a naked 
singularity, (as for instance the negative mass dS Schwarzschild metric), 
or 
\begin{equation} \label{e5.18}
{\mathcal I}^{-} = \emptyset  , 
\end{equation}
i.e. there is no partial conformal completion at past infinity for the 
globally hyperbolic spacetime $(M, g)$. As a very special case, the 
conjecture implies that among all spaces in $dS^{\pm}$, the mass of the 
pure de Sitter space is maximal. In particular, in dimension 4, 
\begin{equation} \label{e5.19}
m_{dS^{\pm}} \leq  0 = m_{0}, 
\end{equation}
where $m_{0}$ is the mass of pure de Sitter. This conjecture has been 
verified in a number of cases [15], [17]; see also [18]-[20] for a 
discussion of possible counterexamples.

 Consider the family of dS Taub-NUT metrics $g^{TN}$ analysed in \S 4. 
For $E$ in the range (4.8), $g^{TN} \in  dS^{\pm}.$ On the other hand, 
(4.4), (4.5), (5.3) and (5.16) show that the mass of $g^{TN}$ is given by
\begin{equation} \label{e5.20}
m(g^{TN}) = -m(g_{TN}) = c\frac{l}{2}E^{1/2}(E-1), 
\end{equation}
for a fixed numerical constant $c$. Hence, when $E > 1$, the mass is 
greater than that of pure de Sitter.

 One might consider strengthening the conjecture to require that 
$\gamma^{+}$ is the round metric $\gamma_{0}$ on $S^{3}$, so that $g$ 
is strongly $dS^{+}$. The conformal Killing field giving the mass is 
thus fixed to be the dilatation field (with south pole to north pole 
flow), as are all the quantities in (5.13)-(5.14) except $T$. However, as 
discussed in \S 3, Friedrich's theorem [12] implies that the data 
$(\gamma_{0}, g_{(3)})$ determine a unique solution $(M, g) \in  dS^{\pm}$, 
for any sufficiently small transverse-traceless form $g_{(3)}$. Since 
$T \sim g_{(3)}$, $T$ is freely specifiable as long as it is small, and 
so from (5.14) it is clear that there exists $\varepsilon_{0} > 0$ such 
that $m$ may assume any value in 
\begin{equation} \label{e5.21}
m \in  (-\varepsilon_{0}, \varepsilon_{0}), 
\end{equation}
within the class of spaces in $dS^{\pm}$. We recall also that in 
dimension 4, the mass is independent of the choice of representative 
$\gamma\in [\gamma]$, so that (5.21) also holds if ${\mathcal I}^{+}$ is 
given by static coordinates, ${\mathcal I}^{+} = {\mathbb R}\times S^{2}$, 
with $\gamma_{0}$ the round product metric.

  There are well-known positive mass theorems in AdS or AH spaces, cf. [49] 
for recent work and references. However, these require that $(M, g)$ is 
strongly AH in that the conformal infinity is the round product metric on 
${\mathbb R}\times S^{n-1}$ or the round metric on $S^{n}$. The only globally 
smooth AH Einstein space with such conformal infinity is hyperbolic $(n+1)$-space, 
by the isometry extension result in [29]; (similar results hold for static 
AdS spaces). Hence, a positive mass result is uninteresting in this context. The 
proper context for the positive mass results is that $(M, g)$ is a strongly AH 
initial data set in a $\Lambda < 0$ vacuum spacetime of one higher dimension, so 
that $(M, g, K)$ satisfies the constraint equations, not the Einstein equations. 
On the other hand, for any boundary metric $\gamma$ on $S^{3}$ or $S^{n}$ 
close to the round metric, there are global AH Einstein metrics with boundary 
metric $\gamma$, cf. [50], and for such spaces, the holographic mass will also 
satisfy (5.21). 

 In general even dimensions, the Taub-NUT metrics on $M = {\mathbb R}\times 
S^{n}$, $n$ odd, within $dS^{\pm}$ again have mass satisfying (5.21), 
and so also violate (5.19). Moreover, one also expects that Friedrich's 
theorem holds in all dimensions, especially in the case where 
$({\mathcal I}^{+}, \gamma) = (S^{n}, \gamma_{0})$, where $\gamma_{0}$ is the 
round metric on $S^{n}$ so that there are no $\log$ terms in the expansion, 
(the variation of the conformal anomaly vanishes). Note also that although 
the mass depends on the choice of representative $\gamma  \in  [\gamma_{0}]$ 
on $S^{n}$, the relations 
$$m > m_{0}, \ \ {\rm or} \ \  m < m_{0}, $$
are independent of this choice; if either inequality holds for one 
choice of $\gamma \in [\gamma_{0}]$, then it holds for all choices in $[\gamma_{0}]$. 

\section*{Appendix}
\setcounter{equation}{0}
\begin{appendix}
\setcounter{section}{1}

 In this Appendix, we carry out the computations verifying (2.17). The 
equation (2.5) states that
\begin{equation} \label{eA.1}
\rho\ddot g_{\rho} -(n-1)\dot g_{\rho} - 2Hg_{\rho} =\rho L, 
\end{equation}
where $L = [2Ric_{\rho} - H\dot g_{\rho} + (\dot g_{\rho})^{2}]$. Also, 
the Gauss-Codazzi equations for $S(\rho) \subset  (M, \bar g)$ and the 
trace of the Riccati equation give
\begin{equation} \label{eA.2}
Ric(T,X) = (\delta A + dH)(X) = 0,
\end{equation}
\begin{equation} \label{eA.3}
H'  + |A|^{2} + 2nR = 0, 
\end{equation}
where $R$ is the scalar curvature of $S(\rho)$.  Equivalently, in local 
geodesic coordinates,
$$\rho \ddot g_{ij} - (n-1)\dot g_{ij} - g^{kl}\dot g_{kl}g_{ij} = 
\rho(2Ric_{ij} - \frac{1}{2}g^{kl}g_{kl}' g_{ij}' + g^{kl}g_{ik}' g_{jl}'),$$
$$g^{kl}(\nabla_{l}\dot g_{ik}  - \nabla_{i}\dot g_{kl} ) = 0,$$
$$\frac{1}{2}g^{kl}\dot g_{kl} + \frac{1}{4}g^{ik}\dot g_{ij} \dot g_{jk} + 2nR = 0.$$
We need to divide into the cases $n$ odd or $n$ even.

 Case I. $n$ odd.

 Suppose first $n = 3$. Differentiating (A.1) once gives
\begin{equation} \label{eA.4}
\rho g_{\rho}^{(3)} - g_{\rho}^{(2)} -2(Hg_{\rho})^{(1)}= (\rho L)^{(1)}. 
\end{equation}
Hence, at $\rho = 0$,
\begin{equation} \label{eA.5}
-g_{(2)} = H'  + Ric_{\gamma} = Ric_{\gamma} - \frac{R_{\gamma}}{4}\gamma . 
\end{equation}
Differentiating again gives
\begin{equation} \label{eA.6}
\rho g_{\rho}^{(4)} - 2(Hg_{\rho})^{(2)} = (\rho L)^{(2)}. 
\end{equation}
The term $Ric_{\rho}$ involves two tangential derivatives of 
$g_{\rho}.$ Since $g^{(1)} = 0$ at $\rho = 0$, the right side of (A.6) 
vanishes at $\rho = 0$. Further, since $(Hg_{\rho})^{(2)} = 
ctrg_{(3)}g$, (A.6) implies $trg_{(3)} = 0$. Here and below $c$ 
denotes a non-zero numerical constant, which may change from line to 
line. The Gauss-Codazzi equation (A.2) then gives $\delta g_{(3)} = 0$. 
However, the transverse-traceless part of $g_{(3)}$ is undetermined by 
(A.6). 

 Differentiating further gives
$$\rho g_{\rho}^{(5)} + g_{\rho}^{(4)} - 2(Hg_{\rho})^{(3)}= (\rho 
L)^{(3)},$$
so that at $\rho  =$ 0, 
\begin{equation} \label{eA.7}
g_{(4)} - trg_{(4)}g_{(0)} + ctrg_{(2)}g_{(2)} = L^{(2)}. 
\end{equation}
This determines the coefficient $g_{(4)}$ in terms of $g_{(2)}$ and its 
2nd tangential derivatives. Taking the next derivative at $\rho  =$ 0 
gives,
\begin{equation} \label{eA.8}
2g_{(5)} - tr g_{(5)}g_{(0)} = L^{(3)}, 
\end{equation}
so that $g_{(5)}$ depends on $g_{(3)}$ and its 2nd tangential 
derivatives. Finally, a further differentiation gives
\begin{equation} \label{eA.9}
3g_{(6)} - tr g_{(6)}g_{(0)} = L^{(4)} + l.o.t., 
\end{equation}
where l.o.t. denotes lower order terms already determined. This 
determines the trace-free part of $g_{(6)}$ in terms of $g_{(4)}$ and 
its 2nd tangential derivatives. The trace of (A.9) is 0. However, 
taking the $\rho$-derivative of (A.3) 4 times also shows that $trg_{(6)}$ 
is determined by $g_{(4)}$ and its 2nd tangential derivatives. Continuing 
this process inductively, one thus sees that the higher coefficients 
$g_{(k)}$ are uniquely determined and depend on the 2nd tangential 
derivatives of $g_{(k-2)}.$

 Consider next the dS side. Here the equation (2.5) is replaced by (2.11), 
giving
\begin{equation} \label{eA.10}
\rho\ddot g_{\rho} -(n-1)\dot g_{\rho} - 2Hg_{\rho} =-\rho L, 
\end{equation}
We assume now
\begin{equation} \label{eA.11}
g_{(0)}^{dS} = g_{(0)}^{AH}. 
\end{equation}
The same arguments as above then give the relation
\begin{equation} \label{eA.12}
g_{(2)}^{dS} = -g_{(2)}^{AH}. 
\end{equation}
As before $g_{(3)}^{dS}$ is transverse-traceless, but is otherwise 
undetermined, and so may be specified arbitrarily. Thus, we choose
\begin{equation} \label{eA.13}
g_{(3)}^{dS} = -g_{(3)}^{AH}. 
\end{equation}
It then follows from (A.10), (A.13) and formulas as in (A.7)-(A.9) that
\begin{equation} \label{eA.14}
g_{(4)}^{dS} = g_{(4)}^{AH}, \ \ g_{(5)}^{dS} = g_{(5)}^{AH}, 
\end{equation}
while
\begin{equation} \label{eA.15}
g_{(6)}^{dS} = -g_{(6)}^{AH}, \ \ g_{(7)}^{dS} = -g_{(7)}^{AH}, 
\end{equation}
and so on. 

 The same analysis and pattern as above holds for arbitrary $n$ odd, 
with $g_{(n)}$ in place of $g_{(3)}$. This verifies (2.17) when $n$ is 
odd.

 Now suppose $n = 4$. Differentiating (A.1) once gives
\begin{equation} \label{eA.16}
\rho g_{\rho}^{(3)} - 2g_{\rho}^{(2)} - 2(Hg_{\rho})^{(1)} = (\rho 
L)^{(1)}, 
\end{equation}
so that at $\rho  =$ 0,
\begin{equation} \label{eA.17}
-g_{(2)} = \frac{1}{2}(Ric_{\gamma} - \frac{R_{\gamma}}{6}\gamma ). 
\end{equation}
Differentiating again gives
\begin{equation} \label{eA.18}
\rho g_{\rho}^{(4)} - g_{\rho}^{(3)}- 2(Hg_{\rho})^{(2)}= (\rho 
L)^{(2)}. 
\end{equation}
As above, the right side of (A.18) vanishes at $\rho  =$ 0. Taking the 
trace of the left-side of (A.18) implies $trg_{(3)} =$ 0, and hence 
$g_{(3)} =$ 0. At the next level, one has
\begin{equation} \label{eA.19}
\rho g_{\rho}^{(5)} - 2(Hg_{\rho})^{(3)}= (\rho L)^{(3)}. 
\end{equation}
Setting $\rho  = 0$, the left side is $ctrg_{(4)} +$ l.o.t. and hence 
$tr g_{(4)}$ is determined by 2nd derivatives of $g_{(2)}$. Similarly 
$\delta g_{(4)}$ is determined via (A.3). However, the right side of 
(A.19) is not pure trace in general. To obtain a consistent expansion, 
one needs to introduce $\log$ terms in the expansion for $g_{\rho}$. Thus, 
set
\begin{equation} \label{eA.20}
g_{\rho} = g_{(0)} + \rho^{2}g_{(2)} + \rho^{4}g_{(4)} + 
\rho^{4}\log\rho \ h_{(4)}. 
\end{equation}
Since $(\rho^{4}\log\rho )^{(5)} = 24\rho^{-1},$ (A.19) gives
\begin{equation} \label{eA.21}
24h -2(Hg_{\rho})^{(3)}= (\rho L)^{(3)}. 
\end{equation}
It follows that $trh_{(4)} = 0$, and by (A.3), $\delta h_{(4)} = 0$ also. The 
equation (A.21) thus determines the transverse-traceless part of $h$, in 
terms of the 2nd tangential derivatives of $g_{(2)}$.

 However, the transverse-traceless part of $g_{(4)}$ is undetermined. 
Next, setting
\begin{equation} \label{eA.22}
g_{\rho} = g_{(0)} + \rho^{2}g_{(2)} + \rho^{4}g_{(4)} + 
\rho^{4}\log\rho \ h_{(4)} + \rho^{5}g_{(5)}, 
\end{equation}
gives
\begin{equation} \label{eA.23}
\rho g_{\rho}^{(6)}+ g_{\rho}^{(5)} - 2(Hg_{\rho})^{(4)}= (\rho 
L)^{(4)}. 
\end{equation}
Observe that $\rho (\rho^{4}\log\rho)^{(6)}+ (\rho^{4}\log\rho)^{(5)} 
= 0$, leaving the term $g_{\rho}^{(5)} - ctr g_{(5)}g_{(0)} +$ l.o.t. 
on the left equal to the right side of (A.23). But the right side 
vanishes, and hence $g_{(5)} = 0$. Thus set 
\begin{equation} \label{eA.24}
g_{\rho} = g_{(0)} + \rho^{2}g_{(2)} + \rho^{4}g_{(4)} + 
\rho^{4}\log\rho \ h_{(4)} + \rho^{6}g_{(6)} + \rho^{6}\log\rho \ h_{(6)}. 
\end{equation}
Differentiating again gives
\begin{equation} \label{eA.25}
\rho g_{\rho}^{(7)}+ 2g_{\rho}^{(6)} - 2(Hg_{\rho})^{(5)}= (\rho L)^{(5)}. 
\end{equation}
For the $\rho^{4}\log\rho h_{(4)}$ term, one has $\rho (\rho^{4}\log\rho)^{(7)} 
+ 2(\rho^{4}\log\rho )^{(6)} = 0$, while for the $\rho^{6}g_{(6)}$ term 
$\rho (\rho^{6}\log\rho)^{(6)}h_{(6)} =  ch_{(6)} + c\log\rho h_{(6)}$. 
Expanding out the right side of (A.25), one finds that $h_{(6)}$ and 
$g_{(6)}$ depend on two tangential derivatives of $g_{(4)}$ and $h_{(4)}$, 
and lower order terms. 

 Continuing in this way inductively shows that the expansion (2.8) is 
even in powers in $\rho$, with $(\log \rho)^{k}$ terms appearing at 
higher orders. Each coefficient $g_{(k)}$ and $h_{(k)}$ depends on two 
derivatives of the lower order terms $g_{(k-2)}$ and $h_{(k-2)}$.

 The dS side when $n = 4$ is analysed exactly as in the case $n = 3$. 
Thus, (A.10) holds and assuming (A.11), (A.12) holds just as before. 
The odd coefficients vanish on both dS and AH sides. Setting
\begin{equation} \label{eA.26}
g_{(4)}^{dS} = g_{(4)}^{AH}, 
\end{equation}
one finds
$$g_{(2k)}^{dS} = (-1)^{k}g_{(2k)}^{AH}, \ \ {\rm and} \ \  
h_{(4)}^{dS} = h_{(4)}^{AH}. $$
The same arguments and patterns hold for any even $n$, and give (2.17). 

\end{appendix}

\bibliographystyle{plain}

\medskip

\begin{center}
July, 2004
\end{center}

\medskip
\noindent
\address{Department of Mathematics\\
SUNY at Stony Brook\\
Stony Brook, NY 11794-3651}

\noindent
\email{anderson@math.sunysb.edu}

\end{document}